\documentclass[a4paper,11pt]{article}
\pdfoutput=1 % if your are submitting a pdflatex (i.e. if you have
\usepackage{jheppub} % for details on the use of the package, please
\usepackage[T1]{fontenc} % if needed

\title{Neutral Exotica at FASER$\nu$ and SND@LHC}

\author{S. Ansarifard and Y. Farzan}
\affiliation{School of physics, Institute for Research in Fundamental Sciences (IPM)
\\
P.O.Box 19395-5531, Tehran, Iran}

\emailAdd{ansarifard@ipm.ir}
\emailAdd{yasaman@theory.ipm.ac.ir}

\abstract{The $(g-2)_\mu$ anomaly indicates that the second generation of leptons should have new interactions beyond the standard model. The high flux of $\nu_\mu$ and $\bar{\nu}_\mu$ at the forward experiments such as FASER$\nu$ and SND@LHC makes them suitable setups to search for new interactions of the second generation leptons. In this paper, we build a model in which the second generation left-handed leptons couple to a new right-handed neutrino, $N$ and a new Higgs doublet which also couples to the quarks. The scattering of high energy $\nu_\mu$ off nuclei can produce $N$. We investigate how forward experiments can test this model by looking for the $N$ production vertex followed by the displaced vertex of the $N$ decay. Discovering even a single such event can be a harbinger to look for the spectacular signals of the new Higgs doublet production at the LHC. We discuss the possibility of explaining the $(g-2)_\mu$ anomaly by adding more generations of $N$ which will lead to chain decays of $N$ and multiple leptons with distinct signals both at forward experiments and at the CMS and ATLAS detectors. Finally, we show that by adding a new light singlet scalar mixed with the neutral component of the new Higgs doublet ({\it i.e.,} 2HDM+$S$ model), the statistics of the data sample can be dramatically increased.}

\begin{document} 
\def\d{{\rm d}}
%%%%%%%%%%%%%%%%%%%%%%%
\def\Epos{E_{\rm pos}}
\def\ap{\approx}
\def\eff{{\rm eft}}
\def\L{{\cal L}}
\newcommand{\vev}[1]{\langle {#1}\rangle}
\newcommand{\CL}   {C.L.}
\newcommand{\dof}  {d.o.f.}
\newcommand{\eVq}  {\text{EA}^2}
\newcommand{\Sol}  {\textsc{sol}}
\newcommand{\SlKm} {\textsc{sol+kam}}
\newcommand{\Atm}  {\textsc{atm}}
\newcommand{\Chooz}{\textsc{chooz}}
\newcommand{\Dms}  {\Delta m^2_\Sol}
\newcommand{\Dma}  {\Delta m^2_\Atm}
\newcommand{\Dcq}  {\Delta\chi^2}
\newcommand{\nbb}{$\beta\beta_{0\nu}$ }
\newcommand {\be}{\begin{equation}}
\newcommand {\ee}{\end{equation}}
\newcommand {\ba}{\begin{eqnarray}}
\newcommand {\ea}{\end{eqnarray}}
\def\VEV#1{\left\langle #1\right\rangle}
\let\vev\VEV
\def\e6{E(6)}
\def\10{SO(10)}
\def\21{SA(2) $\otimes$ U(1) }
\def\321{$\mathrm{SU(3) \otimes SU(2) \otimes U(1)}$ }
\def\lr{SA(2)$_L \otimes$ SA(2)$_R \otimes$ U(1)}
\def\422{SA(4) $\otimes$ SA(2) $\otimes$ SA(2)}

\def\roughly#1{\mathrel{\raise.3ex\hbox{$#1$\kern-.75em
      \lower1ex\hbox{$\sim$}}}} \def\lsim{\roughly<}
\def\gsim{\roughly>}
\def\ltap{\raisebox{-.4ex}{\rlap{$\sim$}} \raisebox{.4ex}{$<$}}
\def\gtap{\raisebox{-.4ex}{\rlap{$\sim$}} \raisebox{.4ex}{$>$}}
\def\lsim{\raise0.3ex\hbox{$\;<$\kern-0.75em\raise-1.1ex\hbox{$\sim\;$}}}
\def\gsim{\raise0.3ex\hbox{$\;>$\kern-0.75em\raise-1.1ex\hbox{$\sim\;$}}}

\maketitle
\flushbottom

\section{Introduction}
During the last 90 years, several breakthroughs in the field of particle physics and high energies have been brought about by studying the properties of neutrinos. This historical fact motivates scrutinizing neutrino interaction at all possible energy intervals to look for signs of new non-standard interactions. BOREXINO has made precise measurement of interaction of solar neutrinos with energies of sub MeV to 14~MeV on the electron. The interaction of neutrinos with energy range of few 10~GeV off nuclei has been studied with great precision by experiments like NOMAD \cite{NOMAD:2001xxt}. Finally, neutrino telescopes such as ICECUBE, DEEPCORE, ANTARES, ARCA and ORCA  detect the scattering of higher energy atmospheric and cosmic neutrinos on nuclei but these detectors cannot resolve the fine details of the scattering processes in case an intermediate new particle is produced.

Neutrinos of all flavors with 100~GeV$-$1~TeV energies can be  produced at the Interaction Points of the LHC. The main detectors of the LHC, CMS and ATLAS, are not designed to detect these neutrinos and miss them.
Indeed, interactions of colliding protons produce large flux of quarks along the beamline which hadronize as pions, Kaons and charmed hadrons. The eventual decay of these hadrons emit a large flux of neutrinos in the forward direction. During the run III of the LHC in 2022-2024, two new detectors called FASER$\nu$ \cite{Abreu:2019yak} and SND@LHC 
\cite{SND} will detect these large fluxes of neutrinos emitted in the forward direction. These detectors are designed to resolve even the short track of the $\tau$-lepton produced by Charged Current (CC) $\nu_\tau$ interaction. That is they have a superb spatial resolution which makes them ideal for probing new feebly interacting particles that  go through chain decays.

Ref.~\cite{Bakhti:2020vfq} demonstrates that FASER$\nu$ can explore a variety of light dark matter models in which dark matter abundance is set via freeze-out scenario with  annihilation into pairs of intermediate light new neutral particles. 
Ref. \cite{Falkowski:2021bkq} studies the capability of FASER$\nu$ to probe effective beyond standard model couplings between neutrino and quarks that may lead to anomalous Charged Current (CC) interaction of neutrinos: $\nu+{\rm nulceus}\to l+X$.
Ref. \cite{Ismail:2020yqc} makes a forecast of   the impact of the Neutral Current (NC) Non-Standard Interaction (NSI) of form $[(V-A)(V \pm A)]$ on FASER$\nu$ data:  $\nu +{\rm nucleous} \to \nu +X$.
Ref.~\cite{Kling:2020iar} investigates the  effects of light flavor gauge bosons at FASER$\nu$ and SND@LHC.
Ref.~\cite{Bakhti:2020szu} as well Ref.~\cite{Jho:2020jfz,Jodlowski:2020vhr} scrutinize  $\nu+{\rm nucleus} \to N+X$ in	which $N$ is a heavy neutrino. In the prvious studies, the intermediate particle  meditating $\nu+{\rm nucleus} \to N+X$ is either a SM photon (interacting via dipole interaction \cite{Jodlowski:2020vhr,Ismail:2021dyp}) or a dark photon or $Z'$ \cite{Bakhti:2020szu,Jho:2020jfz}.

Large fluxes of $\nu_\mu$ and $\bar{\nu}_\mu$ at FASER$\nu$ and at SND@LHC
provide a unique opportunity to study possible interactions of the second generation leptons with new particles. Such new interactions are also motivated by the famous $(g-2)_\mu$ anomaly. In this paper, we build a model in which the second generation left-handed leptons interact with a new right-handed neutrino $N$ and a new scalar $SU(2)$ doublet that also couples to the quarks. $N$ heavier than a few GeV is an uncharted territory which could not be explored by previous neutrino scattering experiments but as we shall show FASER$\nu$ and SND@LHC can explore $N$ with a mass up to $\sim 15$ GeV. We outline the signals of the models at forward experiments and show that the signals are background free. We propose methods to derive the mass and the lifetime of $N$ from the data. We forecast the bounds that can be extracted from the data of  SND@LHC, FASER$\nu$ and its upgrade for HL-LHC on the relevant effective couplings. We also argue that the model can be tested by main detectors of the LHC. Indeed, a single signal event at these forward experiment will be a great motivation to look for signatures of the model at CMS and ATLAS.

We  discuss the contribution to $(g-2)_\mu$ and find that to explain the anomaly, the model should be extended to include more generations of right-handed neutrinos and/or  scalar doublets. Adding more scalar doublets may increase the new effective coupling between $\bar{N}\nu$ and the quarks, increasing the statistics of the signal at the forward experiments. Multiple $N$ can have a more spectacular effect, leading to chain decays of right-handed neutrinos at both forward experiments and at the main detectors of the LHC ({\it i.e.,} at CMS and ATLAS). In the process of chain decay, multiple charged leptons can be emitted. 
Recent studies report such multiple lepton signal in the LHC data \cite{vonBuddenbrock:2017gvy,Fischer:2021sqw,Hernandez:2019geu}.  As discussed in \cite{Sabatta:2019nfg}, 
the $(g-2)_\mu$ anomaly and these multilepton excess at the LHC may be related. 
We have discussed the  signatures of the variant of the model 
with multiple $N$ at forward experiments and discuss how such a variant  can be distinguished from the minimal version of the model with only a single $N$.

 Finally, we add a lighter singlet scalar, $S$, mixed with the neutral component of the scalar electroweak doublet. This makes the model a two Higgs doublet plus $S$ model which is also motivated by some anomalies reported in the LHC data
\cite{vonBuddenbrock:2017gvy,Fischer:2021sqw,Hernandez:2019geu,
	Mathaha:2021buc,Crivellin:2021ubm} (see, however, \cite{Fowlie:2021ldv}). We show that with adding such a light scalar the statistics of the signal at forward experiments can be dramatically increased.

This paper is organized as follows. In sect.~\ref{FE}, we describe the characteristics of the forward experiments that are relevant for our analysis. We then outline the concept of deriving information on the properties of new neutral intermediate particles which might be produced and subsequently decay inside the detector. In sect.~\ref{model}, we propose the minimalistic version of our model and describe its signals at forward experiments. We assess the background for the signal. We compute the cross section of the  $N$ production by $\nu_\mu$ and $\bar{\nu}_\mu$ scattering off nucleons. We forecast  the bounds from forward experiments on the couplings in case of null signals. We also compute maximum number of signal events, saturating the present bounds and discuss how the parameters of the underlying model can be extracted in the lucky situation of relatively large signal sample.   
In sect.~\ref{phen}, we compute the contribution from the new interactions to $(g-2)_\mu$ and to the neutrino masses. We then describe the potential signals at the main detectors of the LHC, CMS and ATLAS.
In sect.~\ref{multiple}, we describe the signatures of a variant of the model with multiple $N$ at forward experiments as well as at the main detectors of the LHC. In sect.~\ref{light-singlet}, we show that by adding  a singlet scalar, the $N$ production cross section can be significantly increased.
In sect.~\ref{summary}, we summarize the results and briefly discuss the implications for the detection of atmospheric neutrinos by neutrino telescopes.

\section{Forward experiments \label{FE}}
In this section, 
we briefly review the setup and capabilities of  SND@LHC, FASER$\nu$ and its upgrades for high luminosity LHC. FASER$\nu$ and SND@LHC will take data during the run III of the LHC. They will be both located at a distance of 480~m from the ATLAS Interaction Point (IP) but at opposite sides in the TI12 and TI18 tunnels \cite{Kling:2021gos}. These detectors are designed to detect neutrinos emitted from the IP in the forward direction. The main sources of neutrinos are the decays of hadrons such as pions, Kaons, and charmed hadrons produced at the IP with a momentum along the beamline. Although the distance of both detectors from the IP are equal, the fluxes of neutrinos at SND@LHC are predicted to be softer than those at the FASER$\nu$ because, while FASER$\nu$ is located exactly in the beamline direction (right before the FASER detector), SND@LHC will be located slightly off-line.
The upgrade of FASER$\nu$ will be located at the proposed Forward Physics Facility (FPF) at a distance of 620~m from IP.
While during run III (2022-2024) 150 fb$^{-1}$ integrated luminosity will be collected at ATLAS IP, during the high luminosity upgrade the integrated luminosity will increase to 3000 fb$^{-1}$. Ref.~\cite{Kling:2021gos} predicts fluxes of neutrinos in these three detectors. 

The FASER$\nu$ and SND@LHC detectors are both made of Tungsten with masses of
 1.2~tons \cite{Abreu:2019yak} and 850~kg \cite{SND}, respectively. Indeed FASER$\nu$ is made of 1000 layers of emulsion films interleaved with Tungsten plates of thickness of 1 mm.  The size of FASER$\nu$ is $25~{\rm cm}\times 25~{\rm cm} \times 1.3~{\rm m}$~\cite{Abreu:2019yak}. 
 That of SND@LHC is $41.6~{\rm cm} \times 38.7~{\rm cm} \times 32~{\rm cm} $. The upgrade of FASER$\nu$ will be of {size $50~{\rm cm}\times 50~{\rm cm} \times 5~{\rm m}$  and} will weight  10~tons~\cite{felix_kling_2020_4059893}.
  
  Notice that FASER$\nu$, being an  emulsion detector,
  cannot record the timing of events. The whole data taking period of run III will be divided into periods with 10-50 fb$^{-1}$ integrated luminosity after which the emulsion will be processed \cite{Abreu:2019yak}. As we will see in sect.~\ref{model}, such division reduces significantly the accidental background from standard model NC and CC events to our signal.
  
The great advantage of these detectors is their spatial resolution which makes them capable of resolving even the short $\tau$-tracks. 
For example, FASER$\nu$ can resolve vertices with a precision of $\sigma_{pos}=0.4~\mu$m. The angular resolution of tracks will be $\sqrt{2} \sigma_{pos}/L_{tr}$ where $L_{tr}$ is the track size. As discussed in  \cite{Bakhti:2020vfq}, this makes FASER$\nu$ ideal setup for studying the dark sector which  go through chain decays. The energy resolution is however modest and of order of 30~\% \cite{Abreu:2019yak}.
 
Let us consider a new particle produced at FASER$\nu$ decaying to $n_f$ final charged particles. The mass of this particle can be  extracted by measuring the invariant mass of these final particles with a relative error of $30 \% \sqrt{n_f}$. If $N_{new}$ such events are observed, the relative precision will be improved to $30\% \sqrt{n_f/N_{new}}$.
The great precision in the angular 
resolution can also help us to find out whether the decay products of the new particle  also include an invisible neutral particle or not. This can be done by reconstructing {transverse missing} momentum relative to the direction of the momentum of the intermediate new particle. Let us consider a new particle (charged or neutral) whose production and decay vertices are both inside the FASER$\nu$ detector and are reconstructed. Let us denote the distance between these two vertices by $L_{new}$. The {transverse momentum} of a possible invisible particle can be reconstructed by projecting the momenta of the final charged particles onto the plane perpendicular to the track of the invisible particles. If the final charged particles make an angle smaller than $90^\circ$ with each other, the {transverse momentum} of the missing light particle will be of order of $m_N/2$. If $m_N/(2E_{new}) > \sqrt{2} \sigma_{pos}/L_{new}$, the precision will be enough to distinguish the emission of a neutral invisible particle in the final state. By measuring $L_{new}$, the lifetime of the new particles can also be extracted. The uncertainty in the deviation of the lifetime  will be mostly dominated by statistics:  $\Delta \tau_{new}/\tau_{new} =2 \log 2/\sqrt{N_{new}}$.

\section{The minimal model and its prediction for forward experiments \label{model}}
In subsect.~\ref{Lag}, we introduce the field content of the minimal model and its Lagrangian. In
subsect.~\ref{Nprod}, we compute the cross section of $\nu_\mu+{\rm nucleon}\to N+X$ and  predict the number of events at the forward experiments. In subsect.~\ref{Ndec}, we study the decay modes  of $N$. We calculate the average energy of $N$ produced by $\nu_\mu$ with a given energy and use it to estimate the decay length of $N$ at the forward experiments. In subsect.~\ref{sig}, we describe the signals and estimate the potential backgrounds for it. We then forecast the bounds that can be derived on the effective coupling by the forward experiments if no signal is observed.
\subsection{The model content and Lagrangian \label{Lag}}

\begin{figure}
\centering
 \includegraphics[width=0.6\textwidth]{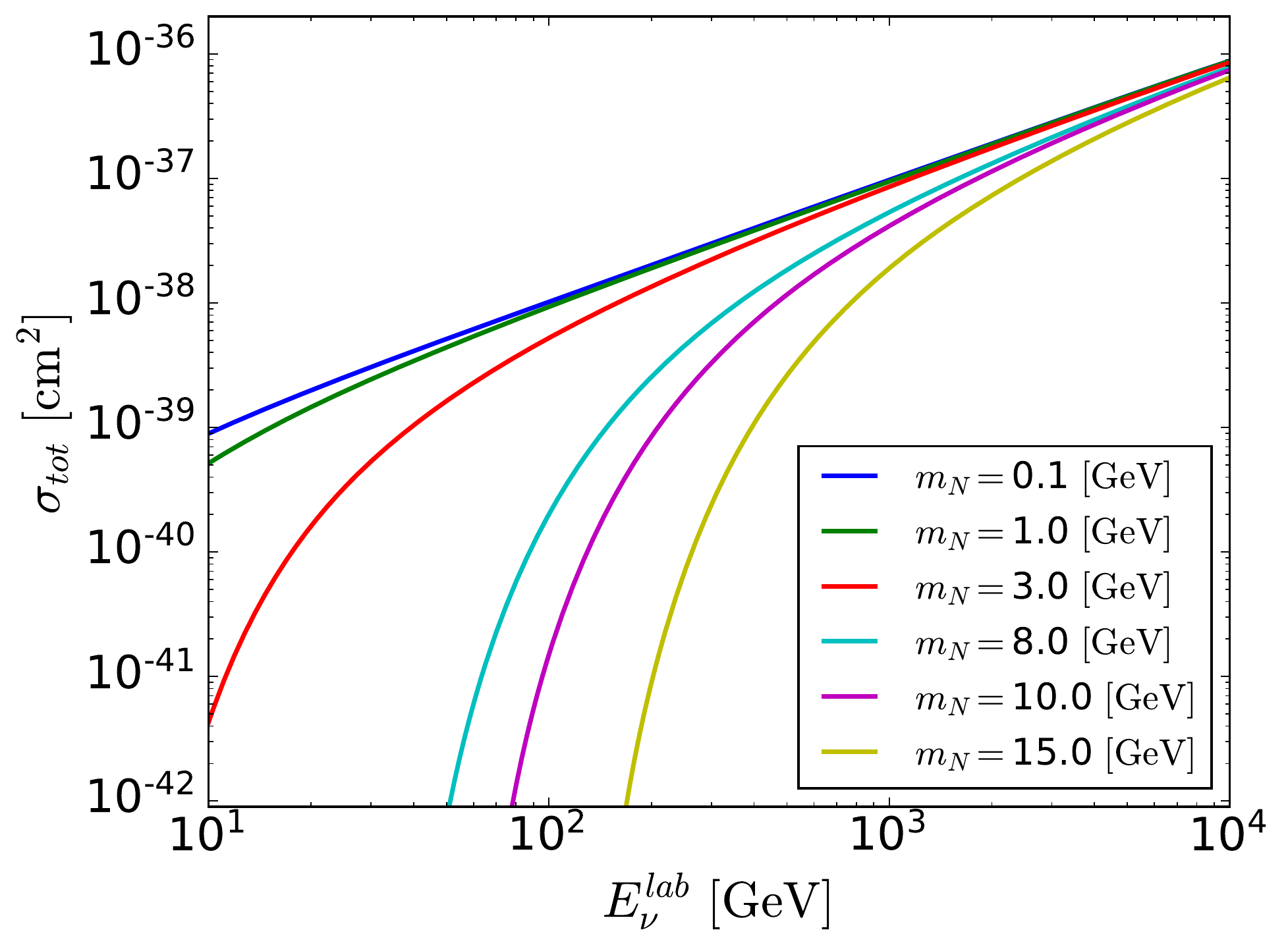}
  \caption{Cross section of $\nu_\mu+{\rm nucleon}\to N+X$ versus the energy of the initial neutrino for various $N$ mass. We have taken $G_u=G_d=10^{-5}$~GeV$^{-2}$. }
  \label{fig:sigma_tot}
\end{figure}

In this subsection, we introduce the Lagrangian of the minimalistic version of the model.
We add a scalar doublet
\be \Phi=\left[ \begin{matrix} \Phi^+ \cr \Phi^0 \end{matrix} \right]\ee 
and a right-handed singlet fermion, $N$ to the standard model. The right-handed $N$ can be either Majorana or Dirac. We turn on the following Yukawa interactions
\be Y_\alpha \bar{N} \Phi^TcL_\alpha+ Y_d \bar{d} \Phi^{\dagger}Q +Y_u \bar{u} \Phi^{T}cQ 
+{\rm H.c.}\ee
where $L_\alpha$ is the left-handed lepton doublet of flavor $\alpha$ and $Q$ is the first generation of left-handed quarks.
As long as the masses of $\Phi^0$ and $\Phi^+$ are heavier than $\sim 300$ GeV and $Y_u,Y_d\stackrel{<}{\sim}0.3$, the present bounds from the direct production of these particles at CMS and ATLAS can be avoided.
Moreover, as long as the splitting between the components is small, the bounds from precision data can be satisfied  \cite{Haller:2018nnx}. 
However, $\Phi^0$ and $\Phi^+$ will be  within the reach of the high 
luminosity phase of the LHC or even that of the run III. We shall discuss the possible  signatures  of direct $\Phi$ production in sect.~\ref{phen}. The Yukawa coupling to the muon, $Y_\mu$ can also lead to a contribution to $(g-2)_\mu$. As we shall see in sect.~\ref{phen}, explaining the famous $(g-2)_\mu$ anomaly motivates large values of $Y_\mu$, saturating the perturbativity condition, $Y_\mu\sim 3-4$. Ref.~\cite{Allwicher:2021rtd} shows that   the models  explaining the $(g-2)_\mu$ anomaly with new Yukawa coupling  generally need large Yukawa couplings. Here, our main aim  is not to explain the $(g-2)_\mu$ anomaly but we take $Y_\mu \sim O(1)$ to have  a high rate of the $N$ production at FASER$\nu$ and SND@LHC detectors due to $\nu_\mu$ interaction. Integrating out $\Phi^0$ and $\Phi^+$, we obtain the following effective Lagrangian
\be G_{u} \bar{N}_R\nu_\mu \bar{u}_Lu_R+ 
G_{d} \bar{N}_R\nu_\mu \bar{d}_Rd_L+
G_{L} \bar{N}_R\mu_L \bar{d}_Ru_L+
G_{R} \bar{N}_R\mu_L \bar{d}_Lu_R+{\rm H.c.}\label{GuGd} \ee
where
$$ G_u=\frac{Y_\mu Y_u^*}{m_{\Phi^0}^2}, \  G_d=-\frac{Y_\mu Y_d}{m_{\Phi^0}^2}, \ 
 G_L=\frac{Y_\mu Y_d}{m_{\Phi^+}^2}, \ {\rm and } \  
 G_R=\frac{Y_\mu Y_u^*}{m_{\Phi^+}^2}. $$
 Taking $Y_u \sim Y_d \sim 0.3$, $Y_\mu \sim 3$ and $m_{\Phi^0}\sim m_{\Phi^+}\sim 300$ GeV, we find $G_u\sim G_d \sim G_L\sim G_R \sim 10^{-5}$ GeV$^{-2}$.

Throughout this paper, we assume that only the second generation leptons couple to $\Phi$ and $N$.  We can also impose the following global $U(1)$ symmetry to set $Y_e=Y_\tau=0$:
\be \Phi \to e^{i \alpha}\Phi, \ \  L_\mu \to e^{-i\alpha}L_\mu, \ \ \mu_R \to e^{-i\alpha}\mu_R, \ \ N\to N, \ \ u\to e^{i\alpha}u, \ \ {\rm and} \ \ d\to e^{-i\alpha}d, \label{glob} \ee
with the rest of SM fields including $Q$, the SM Higgs and the first and third generation leptons being invariant.
Notice that the $u$-quark and $d$-quark masses as well as neutrino mixing break this $U(1)$ symmetry. As a result, this symmetry explains the smallness of the masses of the first generation quarks relative to the rest of quark masses, as a bonus. We shall discuss in sect.~\ref{phen} that this symmetry also prevents a large 1-loop contribution to the neutrino mass in case $N$ is of Majorana type.
\subsection{$N$ production by neutrino scattering on nuclei \label{Nprod}}
Let us consider a neutrino of energy $E_\nu^{lab}$ colliding on a quark which carries a fraction $x$ of the proton momentum.
The $s$ Mandelstam variable of the quark neutrino system will be
\be s=x^2m_p^2+2xm_p E_\nu^{lab} \ee 
where the first term is negligible. The energy of the neutrino and quark in the center of mass is 
\be E_\nu=\left( \frac{xm_p E_\nu^{lab}}{2}.\right)^{1/2}.\ee 
The cross section of scattering off a $u$ quark at the center of mass frame with a scattering angle of $\theta$ can be written as 
\be \frac{d\sigma_u}{d\cos \theta}=\frac{G_u^2}{32 \pi  v_{rel}}
\left( 1-\frac{m_N^2}{4 E_\nu^2}\right)^2\left( \frac{m_{\Phi^0}^2}{t-m_{\Phi^0}^2}\right)^2\left(
E_\nu^2(1-\cos \theta)^2 +\frac{m_N^2}{4}(1-\cos^2\theta)\right)\label{diff-cross}\ee 
where {$m_N$ is the mass of particle $N$}. The Mandelstam variable $t$ can be written as
$$t=2E_\nu(E_\nu-\frac{m_N^2}{4 E_\nu})(\cos\theta -1)\stackrel{<}{\sim} O[(10~{\rm GeV})^2]\ll m_{\Phi^0}^2$$ so the second parenthesis in Eq.~(\ref{diff-cross}) can be approximated by 1. 
The cross section of scattering off $d$-quark is given by a similar formula replacing $G_u$ with $G_d$. Since the mediator is scalar, the cross sections of the scattering off quark and antiquark are equal. Moreover, the cross section of neutrino and antineutrino will be equal. We can therefore write the total cross section of scattering on a nucleon ($i \in \{{\rm proton,~neutron}\})$ as
\be\sigma_{tot}^i=
\int_{-1}^1\int_{x_{min}}^1\left[\frac{d\sigma_{u}}{d \cos \theta}(f_u^i(x,t)+ f_{\bar{u}}^i(x,t))+\frac{d\sigma_{d}}{d \cos \theta}(f_d^i(x,t)+ f_{\bar{d}}^i(x,t))\right] dx d\cos \theta, \label{eq:Tocross} \ee
where $x_{min}=m_N^2/(2 m_p E_\nu^{lab})$.
The nucleon parton distribution functions, $f^i_q(x,t)$, are computed {by LHAPDF6~\cite{buckley2015lhapdf6} using NNPDF3.1~\cite{ball2017parton}.} We have set $G_u=G_d=10^{-5}$ GeV$^{-2}$ throughout this paper. Notice that since we have taken $G_u$ equal to $G_d$, the cross sections on  u-quark and d-quark and as a result on the proton and neutron will be  equal. 
That is, from $G_u=G_d$, we conclude $d\sigma_u/d\cos\theta=d\sigma_d/d\cos\theta$
and therefore $\sigma_{tot}^n=\sigma_{tot}^p$.

	 \begin{figure}[tbp]
	 	\centering
	 	\includegraphics[width=0.5\textwidth]{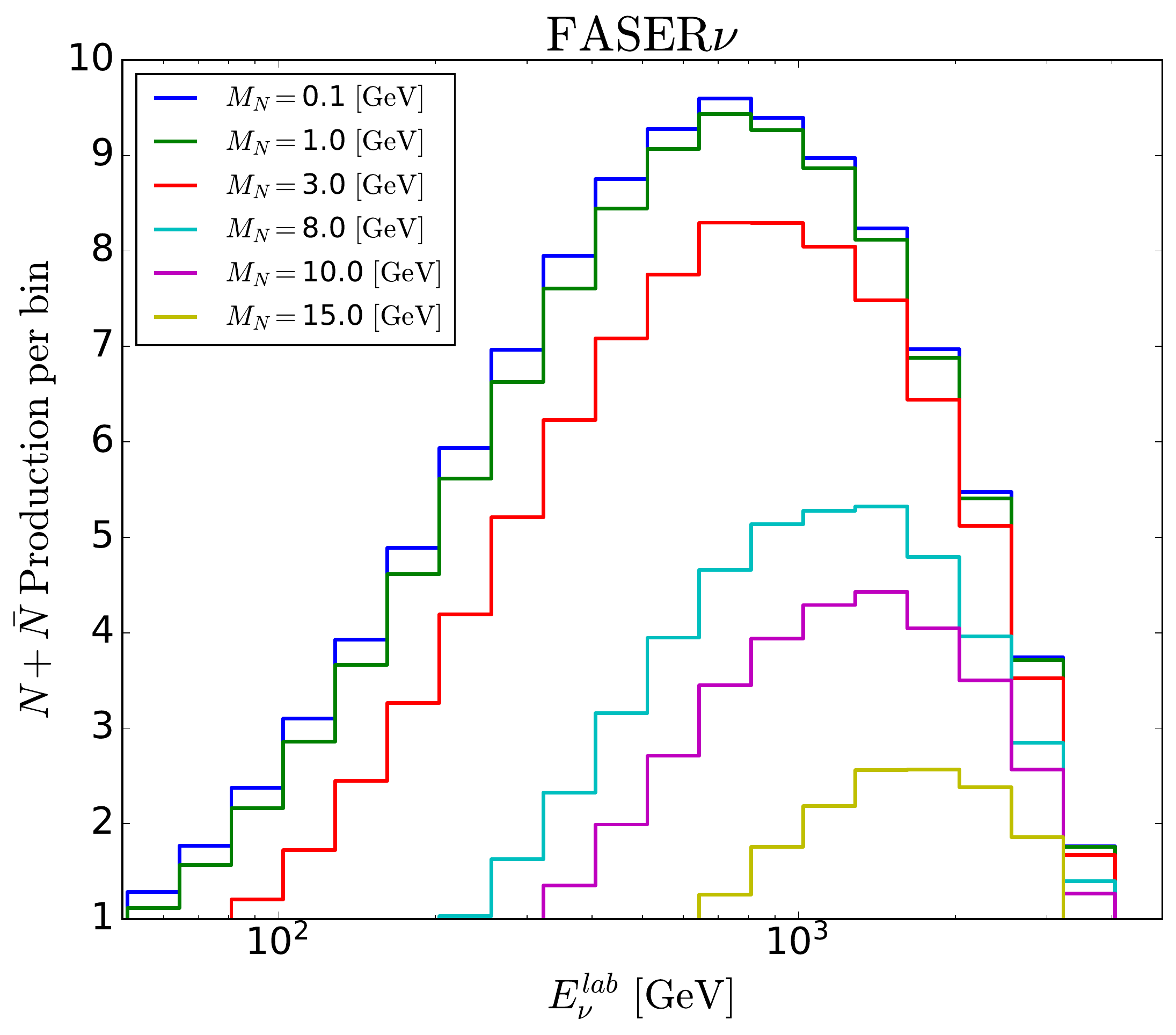}
	 	\caption{The number of $N$ and $\bar{N}$ particles produced by muon neutrino and antineutrino fluxes at each energy bin at FASER$\nu$ during run III of the LHC. We have taken $G_u=G_d=10^{-5}$~GeV$^{-2}$. The fluxes of neutrinos and antineutrinos are taken from \cite{Kling:2021gos}. }
	 	\label{fig:N_tot}
	 \end{figure}

The total cross section  shown in  Eq.~(\ref{eq:Tocross}) versus $E_{\nu}^{lab}$ is illustrated in Fig.~\ref{fig:sigma_tot}, taking $m_N=0.1, 1, 3, 8, 10$~GeV and $15$~GeV. As seen from the figure,    the curves converge for large $E_\nu^{lab}$. This is expected because when the energy of the center of mass is much larger than the masses of final particles, the dependence of the  cross section on the masses becomes weaker. Fig.~\ref{fig:sigma_tot} also shows that for $m_N<1$ GeV and $E_\nu^{lab}>100$~GeV, the cross section is almost independent of $m_N$. This means that the major contribution to the cross section  comes from $x>{\rm few }\times 10^{-2}$. Even in the limit $m_N\to 0$ and taking $G_u=G_d\sim G_F$, the cross section in Eq.~(\ref{eq:Tocross}) is about one order of magnitude smaller than the cross section of  SM neutral current. This is partly due to the spinorial difference of the amplitudes and partly due to the difference in the numerical factors in the definitions of $G_F$ and $G_u$ or $G_d$ (cf., the definitions of $G_u$ and $G_d$ in Eq.~(\ref{GuGd}) with the Fermi interaction, $(4 G_F/\sqrt{2})(\bar{d}_L\gamma^\mu u_L)(\bar{\nu}_\mu \gamma_\mu \mu_L)$).

	 \begin {table}
	 \caption {The total predicted numbers of $N+\bar{N}$ to be detected   by FASER$\nu$ and SND@LHC during run III of the LHC as well as by FASER$\nu 2$ during HL-LHC era, taking $G_u=G_d=10^{-5}$ GeV$^{-2}$ and different values of $m_N$ as shown in the second row. } \label{tab:1} 
	 \begin{center}
	 	\begin{tabular} {| l | c c c c c c|}
	 		\hline 
	 		Number    &    &      &   $N+\stackrel{-}{N} $  &     & &  \\
	 		\hline
	 		$m_N$~GeV      &  0.1  &  1    &   3   &   8   & 10  & 15   \\
	 		\hline
	 		SND@LHC     & 19   &  18    &13   &5   &3   & 1               \\
	 		FASER$\nu$     & 113  &  109  &90 &46  &35 &17               \\
	 		FASER$\nu 2$   & 7685&7394&6045&3019&2229&  1015     \\
	 		
	 		\hline
	 	\end{tabular}       
	 \end{center}
	\end{table}

The total number of $N$ and $\bar{N}$ produced at the detector will be \be
\frac{M_{det}}{m_p} \int_{m_N^2/(2m_p )}(F_{\nu_\mu}(E_\nu^{lab})+F_{\bar{\nu}_\mu}(E_\nu^{lab}))
(r_n	\sigma_{tot}^n+r_p	\sigma_{tot}^p)  d E_\nu^{lab}\ee
where $M_{det}$ is the detector mass and $r_n$ and $r_p$ are  fractions of the neutron and proton in the nuclei of the detector, respectively: $r_p=Z/A$ and $r_n=(A-Z)/A$. 
 $F_{\nu_\mu}$ and $F_{\bar{\nu}_\mu}$  \cite{Kling:2021gos} are time integrated fluxes per unit area at the detector. Fig. (\ref{fig:N_tot}) illustrates the spectrum of $N+\bar{N}$ produced during the run III of the LHC at FASER$\nu$. {The total number of events at SND@LHC, FASER$\nu$ and FASER$\nu 2$ for various values of $m_N$ are shown in table~\ref{tab:1} }
  
%Fig. (\ref{fig:N_tot}) shows the  spectrum of $N+\bar{N}$ produced during the run III of the LHC at FASER$\nu$, at SND@LHC and at FASER$\nu 2$. The masses of FASER$\nu$ and SND@LHC detectors are equal to 1.3~tons \cite{Abreu:2019yak} and 850~kg \cite{SND}, respectively. We also forecast for FASER$\nu 2$ like detectors with a mass of 10~tons. The total number of events for various values of $m_N$ are shown in table \ref{tab:1}.  

 \subsection{$N$ decay \label{Ndec}}
 The produced $N$ travels a distance of $l\sim \Gamma_N^{-1} E_N/m_N$ where
 $\Gamma_N$ is the total decay rate of $N$. Taking $N$ heavier than $\sim 3$~GeV, the following decay modes are available for the 
 $N$ decay
 \be \Gamma(N\to \nu_\mu u\bar{u})=\frac{|G_u|^2}{|G_d|^2}\Gamma(N\to \nu_\mu d\bar{d})=\frac{|G_u|^2}{|G_L|^2+|G_R|^2}\Gamma(N\to \mu u\bar{d}) =\frac{G_u^2 m_N^5}{1024\pi^3} \label{decay}\ee
 where we have neglected the masses of the final particles. Notice that the factor of three that usually appears in denominator from phase space integration in the three body decay modes has been canceled out from Eq. (\ref{decay}) by the factor of three in the numerator from summation on the color of the final quarks. Moreover, we have assumed that $N$ is heavy enough so that the meson resonances are not relevant for the $N$ decay. 
 %For $m_\pi <m_N< O(3)$~GeV, similarly to the $\tau$ lepton decay, $\pi$ or $\rho$ resonance production would dominate the $\tau$ decay so we could not compute the decay rate by assuming the final quarks as free particles. 
 Of course, if $N$ is of Majorana type, it can also decay into the charged conjugates of the above final particles.

\begin{figure}[tbp]
\centering
 \includegraphics[width=0.5\textwidth]{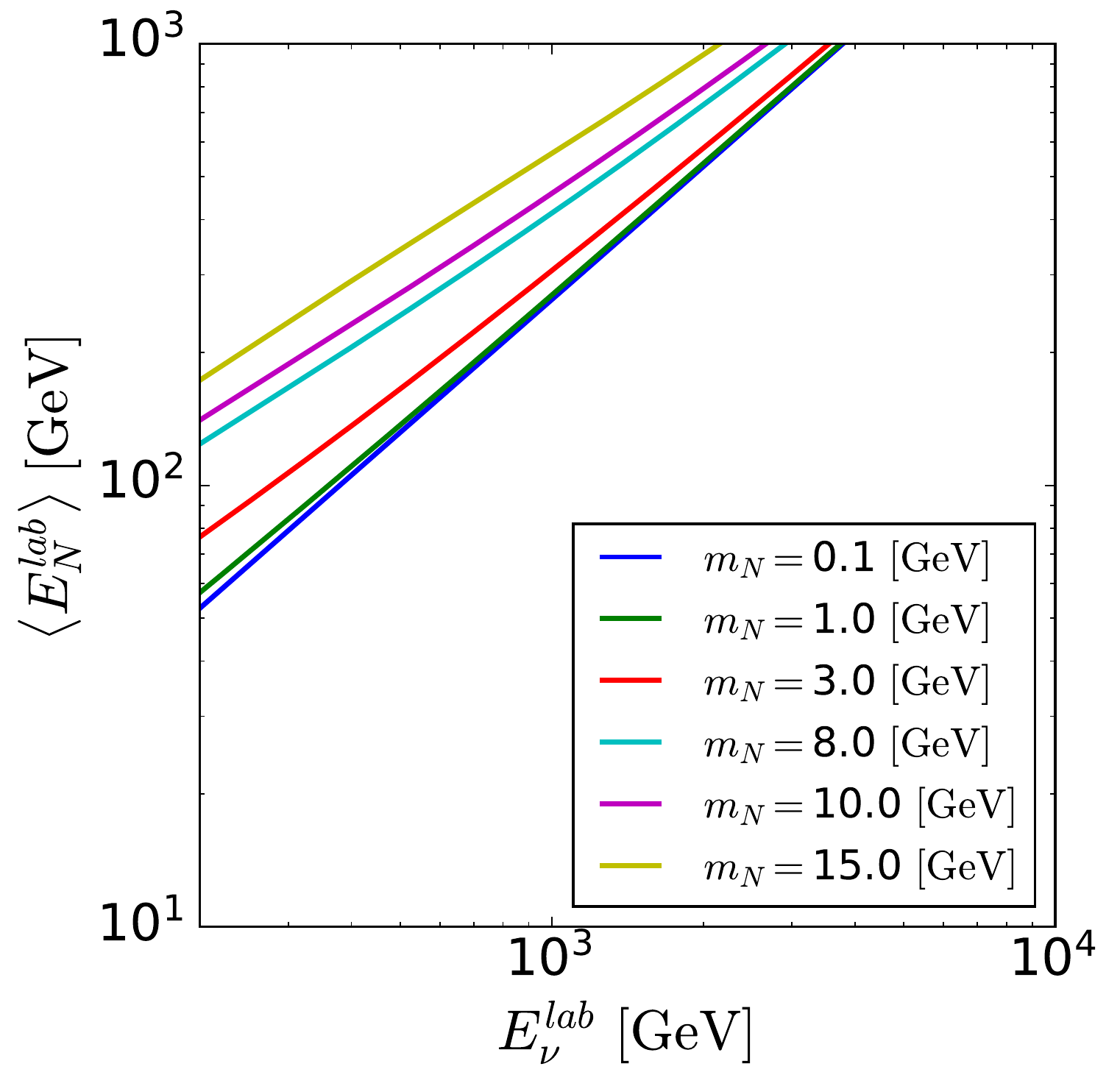}
  \caption{Average energy of $N$ particles  produced by  incoming neutrinos with energy of $E_\nu^{lab}$.   }
  \label{fig:E_N}
\end{figure}

 The average energy of $N$ produced by a neutrino of $E_\nu^{lab}$ can be estimated as 
 \be \langle E_N^{lab}\rangle =\frac{
 \int_{-1}^1\int_{x_{min}}^1 E_N^{lab}\left[\frac{d\sigma_{u}}{d \cos \theta}(f_u(x,t)+ f_{\bar{u}}(x,t))+\frac{d\sigma_{d}}{d \cos \theta}(f_d(x,t)+ f_{\bar{d}}(x,t))\right] dx d\cos \theta}{\sigma_{tot}}, \ee
 where $E_N^{lab}$ in the integrand can be written as 
 $$ E_N^{lab}= \frac{E_\nu^{lab}}{2}(1+\cos\theta) +
 \frac{m_N^2}{4 x m_p}(1-\cos\theta).$$
 $\langle E_N^{lab}\rangle$ versus { $E_{\nu}^{lab}$} is shown in Fig (\ref{fig:E_N}).  As expected $N$ carries a fraction of $O(0.3)$ of the energy of the initial neutrino. For heavier $N$, this fraction is larger because the energy available for the jets produced along with $N$ is lower.
 
 The momentum of $N$ produced by the scattering on a parton of associated momentum fraction $x$  will make an angle of $\sim \gamma^{-1}=(2xm_p/E_\nu^{lab})^{1/2}\sim 10^{-2}$ with the beamline. It will decay after traveling a distance of 
 \be l=\Gamma_N^{-1} \gamma_N=3~\mu{\rm m}\frac{\left(10^{-5}~{\rm GeV}\right)^2 }{|G_u|^2+ |G_d|^2+|G_L|^2+|G_R|^2}\left(\frac{10~{\rm GeV}}{m_N}\right)^6 \left(\frac{E_N^{lab}}{200~{\rm GeV}}\right).\label{decayLENGHT} \ee

\subsection{Signal and background at forward experiments \label{sig}}
\begin{figure}
\centering
 \includegraphics[width=0.8\textwidth]{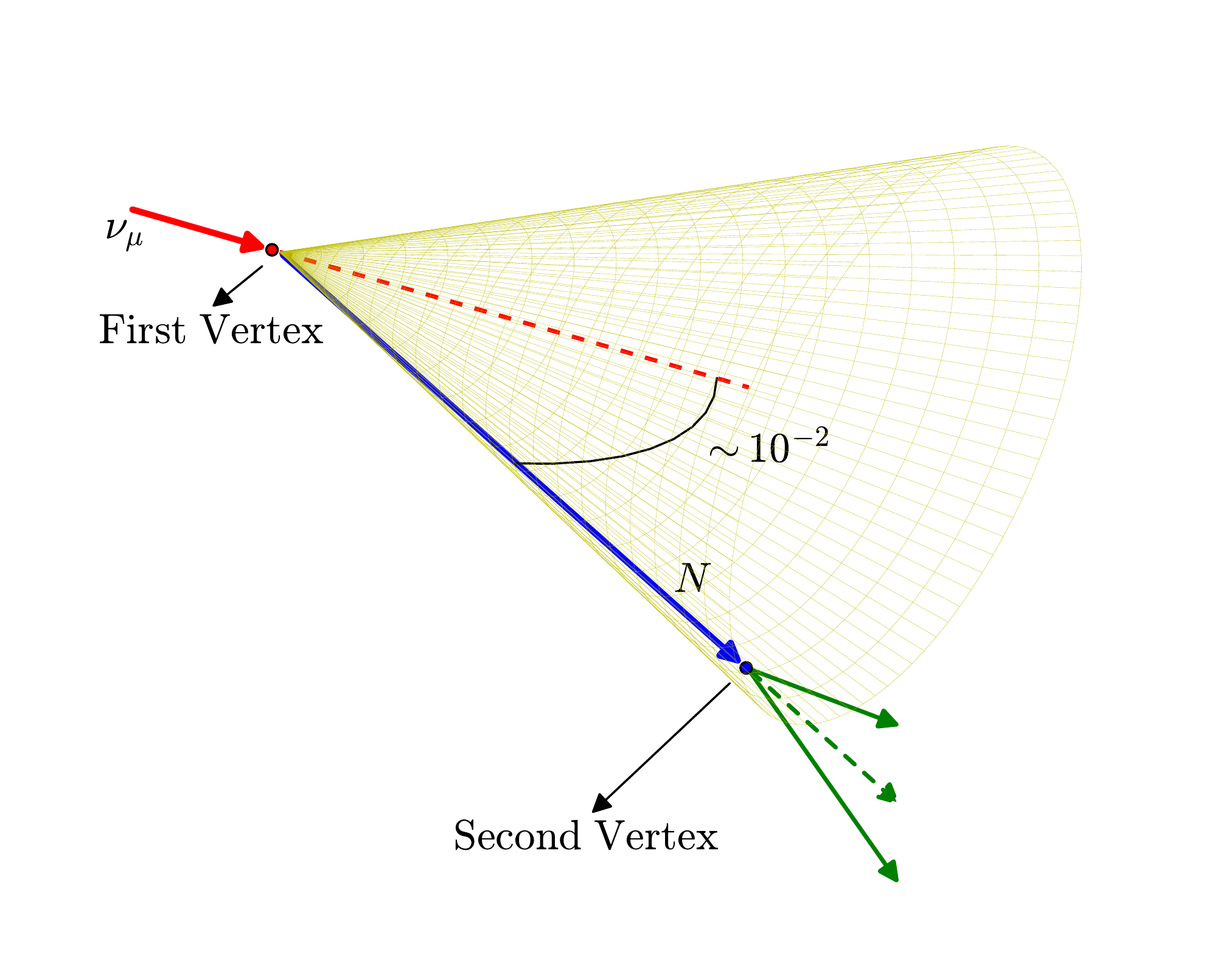}
  \caption{Schematic topology of the signal event. The  green arrows show the final particles from the $N$ decay. The solid green arrows show jets. The dashed green arrow denotes lepton. }
  \label{fig:topology}
\end{figure}

The signature of the production of $N$ vertex is jets similar to the SM neutral current neutrino interaction. As seen from Eq.(\ref{decayLENGHT}), the dependence of the $N$ decay vertex displacement on $m_N$ is very strong. For values of the effective coupling saturating the upper limit, up to $m_N<14$~GeV, the displacement can be large enough to disentangle thanks to the superb position resolution of 0.4~$\mu$m of FASER$\nu$. Taking $m_N>2$ GeV, the displacement will be smaller than 10~cm and therefore within the size of the detector. Since $N$ is also relativistic, its decay products will be emitted within a cone with an opening angle of $m_N/E_N \sim 10^{-3}-10^{-2}$. Again thanks to the excellent angular resolution of FASER$\nu$, it can resolve the two jets associated with $u \bar{u}$ (or $d \bar{d}$) as well as  the two jets associated with $u \bar{d}$ and the charged lepton. The topology of the events are schematically shown in Fig.~\ref{fig:topology}. 
 
The model therefore predicts two signals:  
\begin{itemize} 
\item A neutral-current-like event associated with the $N$ production plus another vertex with two jets associated with $N \to \nu_\mu u \bar{u}$ or $\nu_\mu d \bar{d}$. The second vertex lies within a cone with an apex  at the $N$ production vertex and an opening angle of $\gamma^{-1}\sim 10^{-2}$. The angular separation of the two final jets from each other as well as from the  $N$ track ({\it i.e.,}  from the line connecting the two vertices)    will be also of order of $10^{-2}$. The sum of the momentum of the two jets projected onto the plane perpendicular to the direction of the $N$ momentum ({\it i.e.,} to the direction of the line connecting the $N$ production and the $N$ decay vertices) will be of order of a fraction of $m_N$ which should be compensated with the transverse momentum of the final neutrino which escapes detection.  In almost half of the events, the transverse components of the jet momenta make an angle smaller than $90^\circ$ with each other. For such events, the transverse missing momentum will be relatively large and close to $m_N/2$. As discussed in sect.~\ref{FE}, if the decay length  of the intermediate  $N$ is larger than $\sqrt{2}\sigma_{pos}(2E_{N}/m_N) \sim 20$~$\mu$m, the missing transverse momentum can be reconstructed with enough precision to verify the emission of an invisible particle along with the jets.
\item Again a neutral-current-like event associated with the $N$ production followed by  a  vertex of two jets  plus a muon track  (associated with $N \to \mu u \bar{d}$). The second vertex lies  within a cone with an apex at the $N$ production vertex and an opening angle of $\gamma^{-1}\sim 10^{-2}$. In this case there should not be any missing transverse momentum. By measurement of the four-momenta of the decay products of $N$ and computing their invariant mass, the mass of $N$ can be  reconstructed. By measuring their total momenta the energy of $N$ can be derived. Then, if the statistics is enough, the average displacement of the  vertices (average $l$) gives the lifetime of $N$. The relative error in the extracting the mass and the lifetime of $N$ from $N_{\mu u\bar{d}}$ number of  such types of events will respectively be $40 \%/\sqrt{N_{\mu u\bar{d}}}$ and $2\log 2 /\sqrt{N_{\mu u\bar{d}}}$ as shown in Ref.~\cite{Bakhti:2020vfq}.
\end{itemize}

\begin{figure}[tbp]
	\centering
	\includegraphics[width=0.8\textwidth]{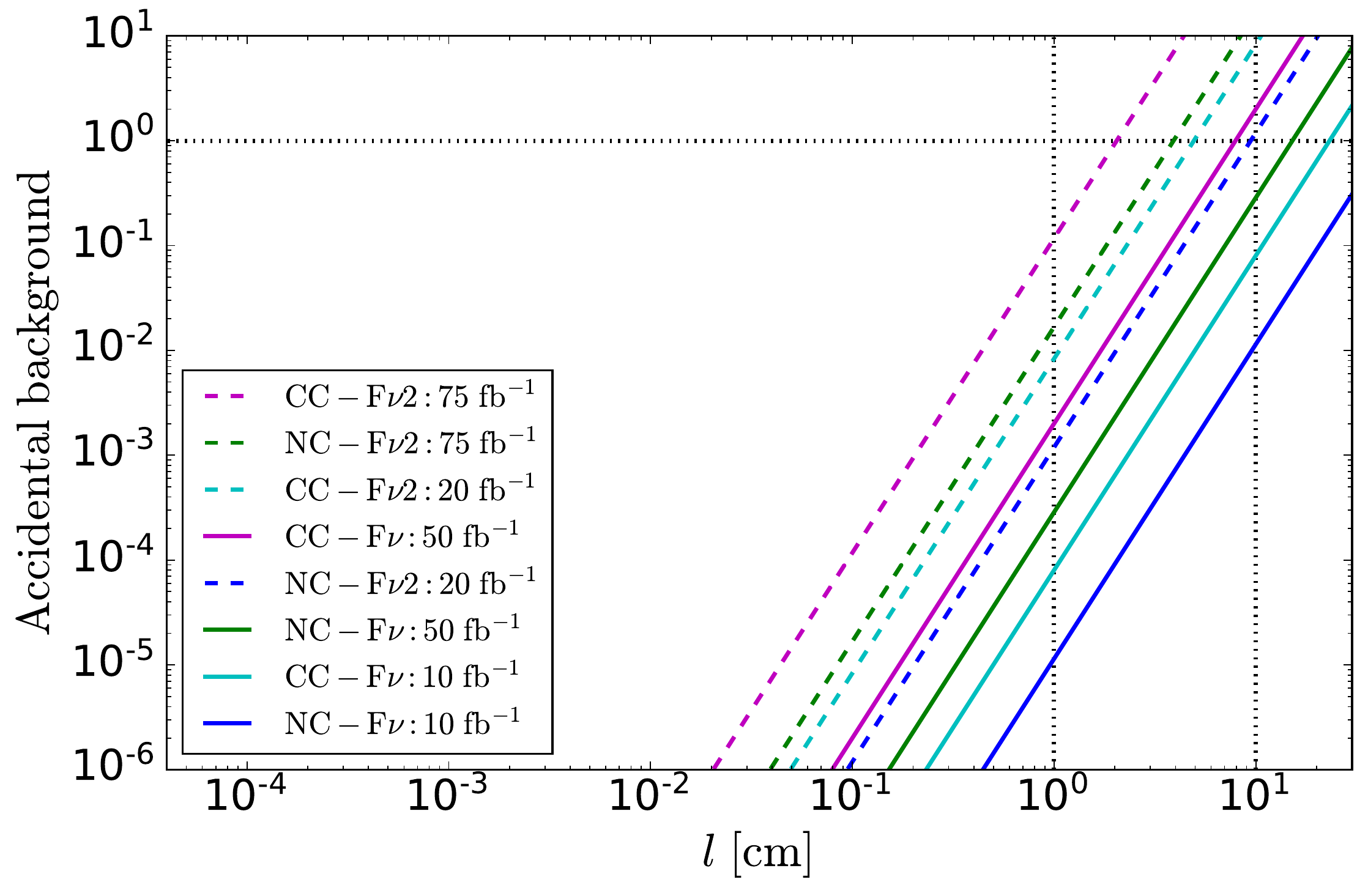}
	\caption{Number of accidental backgrounds for the $N\to \nu q \bar{q}$ and $N\to \mu q' \bar{q}$ signals at FASER$\nu$ and FASER$\nu$2 for data taking periods with collected data of 10, 20, 50 and 75 fb$^{-1}$. Background consists of a NC or CC vertex (respectively, marked as NC or CC) located within a narrow cone of opening angle of $10^{-2}$ and a height of $l$ in front of another neutral current vertex with a topology and orientation as depicted in Fig.~\ref{fig:topology}. The solid and dashed lines respectively correspond to FASER$\nu$ and FASER$\nu$2.  }
	\label{fig:Acc-Back}
\end{figure}

 Let us now estimate the backgrounds for the signals, starting with the main source of the background which is accidental alignment of two SM neutrino scattering vertices. If two separate neutral current events accidentally happen along the beam direction from each other in one data collecting period, they can mimic the first signal described above when processing the emulsion. Similarly, if a $\nu_\mu$ charged current event lies ahead of a neutral current event, it can mimic the second signal described above.
 The volume of a cone with a height of $l$ and an opening angle
  of $\gamma^{-1}$ is $\pi l^3 \gamma^{-2}/3$. The probability of one event to lie accidentally within a cone with the apex at the other vertex  is $p = (\pi l^3 \gamma^{-2}/3)/({\rm detector~ size}).$  The numbers of  fake signals  from pile-up of the SM vertices   are therefore    equal to $\mathcal{N}_{NC}^2\times p/2 $ and $\mathcal{N}_{NC}\times \mathcal{N}_\mu \times p $ where $\mathcal{N}_{NC}$ and $\mathcal{N}_{\mu}$ are respectively the numbers of  SM NC and $\nu_\mu$ CC vertices   passing the applied cuts in each data collecting period. Of course,  for heavier $N$, $l$ is smaller and the relevant probabilities decrease very fast with a factor of $l^3 \propto (2~{\rm GeV}/m_N)^{18}$.
 Fig.~\ref{fig:Acc-Back} shows the numbers of accidental backgrounds for $N\to \nu q \bar{q}$ (marked with NC) as well as for  $N\to \mu q' \bar{q}$ (marked with CC) at FASER$\nu$ and at FASER$\nu$2 versus the displacement of the second vertex, $l$. The second vertex is supposed to  be  located within a cone with an apex at the first vertex and an opening angle of $10^{-2}$, oriented in the forward direction as depicted in Fig.~\ref{fig:topology}.   We have assumed data accumulation periods of 10, 20, 50 and 75   fb$^{-1}$. To draw the figures we have used the results of \cite{Ismail:2020yqc,Kling:2021gos} to estimate $\mathcal{N}_{NC}$ and $\mathcal{N}_{CC}$. As seen from the figure, with $l<$few cm, the accidental background will be negligible.  Notice that for the $m_N$ range of our interest, $l<$few cm corresponds to $G>{\rm few}\times 10^{-7}$~GeV$^{-2}$ which  fortunately coincides with the coupling range giving rise to  fairly significant  number of events at the forward experiments.

We now discuss other possible sources of the background. Neutral hadrons such as neutrons produced in the neutral current events may mimic the signal
associated with $N$ decaying into $\nu q \bar{q}$. The number of such events is not calculated but is expected to be much smaller than muon induced neutral hadrons \cite{Ismail:2020yqc}. Similarly to discriminating between NC events and neutral hadron induced events \cite{Ismail:2020yqc}, our signal can be distinguished from this background by measuring the transverse momentum of  visible final tracks. The interaction length  of neutral hadrons is of order of 10~cm. Inserting a cut of few cm on the displacement of the second vertex  as well as using the criteria of selecting NC events (described in \cite{Ismail:2020yqc}) can eliminate this background. Another potential source of background is displaced vertex caused by the $\tau$ production by $\nu_\tau$  and its subsequent decay at the detector.   
 By measuring the total charge of the $N$ decay products, $N$ can be discriminated against the $\tau$ lepton produced by the charged current interaction of $\nu_\tau$. If the decay length of $N$ is of order of 100-500 $\mu$m, it may be mistaken with neutral $D$-meson decaying into $\mu$. However, $D$-mesons  are mostly  produced by charged current interaction of $\nu_\mu$ off the strange sea quarks inside the proton and neutron. That is the signature of $D$-meson at FASER$\nu$ is a charged current vertex followed by a displaced vertex. Since in our model the first vertex is neutral current, $D$-meson production will not constitute a background to our  signals. In principle, $D$-meson can also be produced via neutral current interaction of neutrinos on intrinsic $c$ quark in the nucleons but the rate will be too low to cause a significant  background to our signal.  We therefore conclude that our signals with $l<$few cm are practically background-free. As a result, detection of even a single event will be an indication of new physics..

\begin{figure}[tbp]
\centering
 \includegraphics[width=0.5\textwidth]{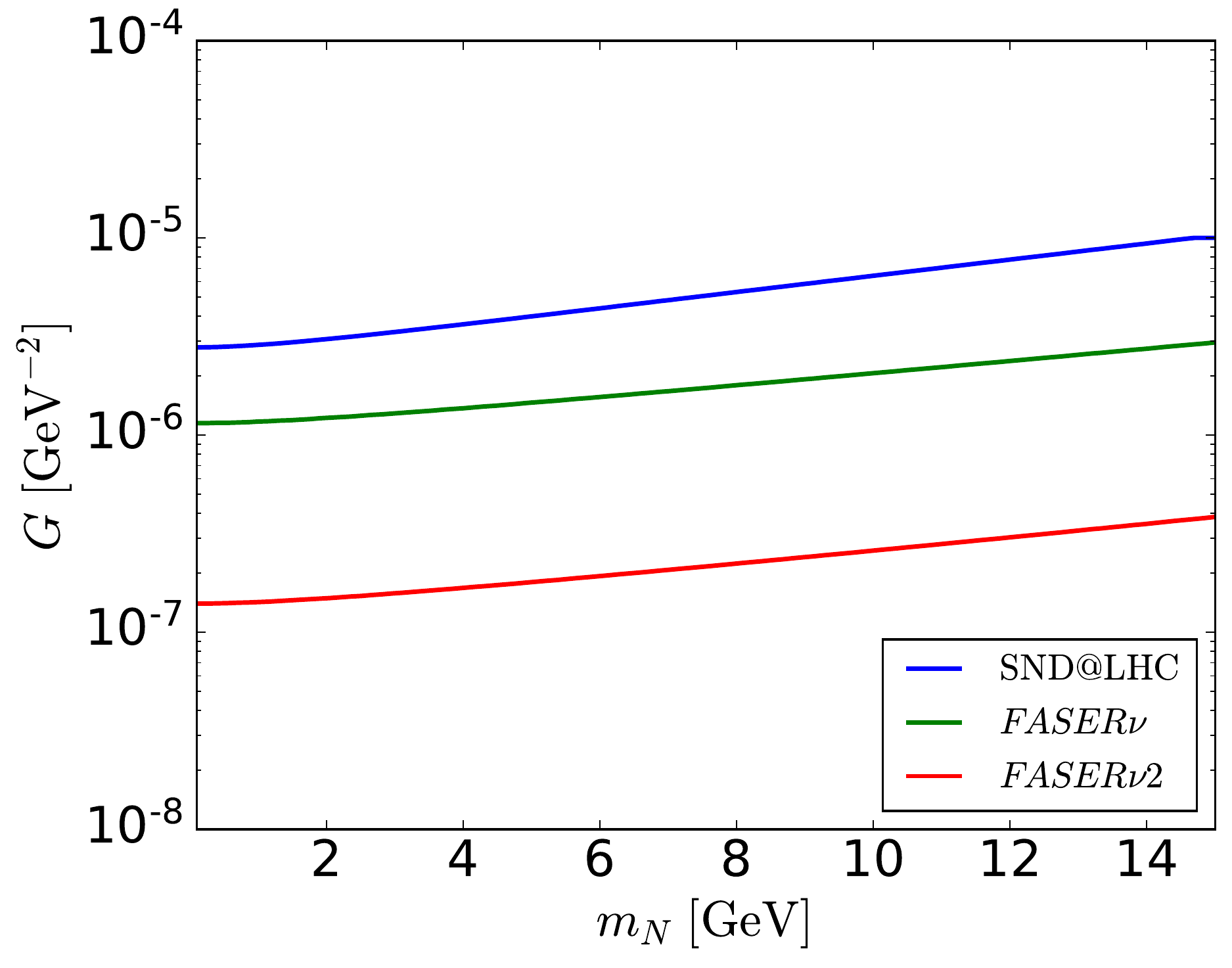}
  \caption{Forecasted upper bound on $G_u=G_d$ from SND@LHC, FASER$\nu$ and its upgrade FASER$\nu$2 in case of null signal. We have assumed integrated luminosity of 150~fb$^{-1}$ and 3000~fb$^{-1}$, respectively, for run III of the LHC (SND@LHC and FASER$\nu$) and for HL-LHC (FASER$\nu$2).  }
  \label{fig:bond}
\end{figure}

Fig.~\ref{fig:bond} shows the bound that  can be set on $G_u=G_d$ versus $m_N$ by FASER$\nu$ and  SND@LHC during the run III of the LHC as well as by the upgrade of FASER$\nu$  during the high luminosity  run of the LHC. That is the curves correspond to number of  produced $N$ equal to 1. As seen from the figure in case of null results, FASER$\nu$ can set a bound of $10^{-6}$~GeV$^{-2}$ on $G_u=G_d$ improving the theoretical limit by a factor of 10. Its upgrade can improve by another order of magnitude. As shown in the figure, for $m_N<15$ GeV, the dependence of the bound from forward experiments on $m_N$ is mild.   If $N$ is lighter than 3~GeV, it should have been already discovered by lower energy neutrino scattering experiments such as NOMAD \cite{NOMAD:2001xxt}.   This is yet another reason why we  focus on $m_N>3$~GeV. 

\section{Neutrino mass, $(g-2)_\mu$ and production of the heavy states at the LHC\label{phen}}
As mentioned before, the $Y_\mu$ coupling gives a contribution to $(g-2)_\mu$ which can be written as \cite{Lavoura:2003xp,Farzan:2009ji}
$$\Delta a_\mu=\delta \left(\frac{g-2}{2}\right)=\frac{Y_\mu^2}{16\pi^2}\frac{m_\mu^2}{m_{\Phi^+}^2}K(m_N^2/m_{\Phi^+}^2),
 $$
where 
 $$K(t)=\frac{2t^2+5t-1}{12(t-1)^3}-\frac{t^2 \log t}{2(t-1)^4}.$$
 For $m_N^2/m_{\Phi^+}^2\to 0$, $K\to 1/12$ so
 $$\Delta a_\mu= \delta \left(\frac{g-2}{2}\right)=5 \times 10^{-10} \left(\frac{Y_\mu}{3}\right)^2 \left(\frac{300~{\rm GeV}}{m_{\Phi^+}^2}\right)^2 .$$
This contribution has the right sign to explain the anomaly but its magnitude is too small to explain the deviation of the experimental result from the SM prediction \cite{Muong-2:2021ojo}. If we want to explain the   $(g-2)_\mu$ anomaly, we have to add more $\Phi$ or $N$ to the model. For example, four $\Phi$ with couplings of $O(3)$ to the muon and $N$ can account for the deviation.
 In this case, the effective couplings can increase by a factor of $\simeq 4$ {leading} to a $O(16)$ fold increase in the statistics at forward experiments such as FASER$\nu$. In case that model has more than one $N$ with masses smaller than $O(10~{\rm GeV})$ and with couplings of $O(3)$ to muon and $\Phi$, the signal of the model at forward experiments as well as at CMS and ATLAS will be completely different. We  shall discuss the phenomenology of such a variant of the model with multiple $N$ in sect. \ref{multiple}. Let us now focus on the consequences of the minimal version with single $N$ and $\Phi$.
 
In our model, the new scalar does not develop a VEV so neutrinos do not obtain a Dirac mass term at the tree level.  The neutrino mass can receive a contribution at one loop level provided that (i) $N$ is a Majorana fermion and (ii) there is a splitting between real (CP-even) and imaginary (CP-odd) components of $\Phi^0$ which  originates from a quartic coupling of form
 $\lambda_{H \Phi} (H^\dagger\cdot \Phi)^2+H.c.$
 \cite{Farzan:2009ji,Boehm:2006mi}. Thus, if we take $N$ to be of Majorana type, the smallness of neutrino 
 mass constrains the splitting of the real and complex components of $\Phi^0$ or  equivalently $\lambda_{H\Phi}$. The smallness of $\lambda_{H\Phi}$ can be explained by the  global $U(1)$ symmetry introduced in Eq.~(\ref{glob}).
    Such a symmetry not only explains the smallness of $\lambda_{H \Phi}$ and the splitting but also explains the smallness of the $u$ and $d$ masses in comparison to those of higher generation quarks as mentioned before.
 
 At the interaction points of the LHC, the components of $\Phi$ can be pair produced via electroweak interaction with a cross section of $\sim 10$ fb \cite{Farzan:2010fw}. The $\Phi$ components can also be singly produced in association of a recoiling gluon via the $Y_u$ and $Y_d$ couplings. Again the cross section is expected to be of order of 10~fb as $\alpha_s Y_u^2, \alpha_s  Y_d^2\sim e^2 \alpha/\sin^4\theta_W$. Since $Y_\mu \gg Y_u,Y_d,e/\sin\theta_W$, the dominant decay modes are $\Phi^0 \to N \bar{\nu}_\mu$ and $\Phi^+ \to N \mu^+$. Subsequently, $N$ will decay to $\mu^-+$two jets or  $\nu_\mu+$two jets.  If $N$ is of Majorana type, it can also decay into the charged conjugates of these final states.
 
  Let us shortly discuss two possible signals: 
 \begin{itemize}
 	\item  $\Phi^+$ in association of a gluon showing up as $\mu^+$ with displaced vertex of $\mu^-+$two jets. If $N$ is of Majorana type, it can even decay into $\mu^++$two jets, providing a distinctive same sign signature. 
 	\item $\Phi^+$ in association of a gluon showing up as $\mu^+$ with displaced vertex of two jets+missing energy.
 \end{itemize}
 Similarly, $\Phi^0$ can be produced in association with the gluon. Moreover, the pair productions $\Phi^+\Phi^-,\Phi^+\bar{\Phi}^0, \Phi^-{\Phi}^0,\Phi^0\bar{\Phi}^0$ and subsequent decays can take place. Exploring all the signals at CMS and ATLAS and assessing their background is beyond the scope of the present paper. However, this rich phenomenology with its distinctive signatures sounds very promising. If FASER$\nu$ finds signals for this model, it will be a great motivation for a dedicated search in the CMS and ATLAS data for the distinctive predictions of the present model. 
 
Notice that FASER$\nu$ cannot distinguish the nature of $N$ ({\it i.e.,} Majorana vs Dirac). The reason is that there is no way to know whether the detected $\mu^-$ is initiated by $\nu_\mu$ via lepton number conserving processes or  by   $\bar{\nu}_\mu$ via lepton number violating processes involving Majorana $N$.
However, at CMS and ATLAS, same sign muon signals coming from $\Phi^+\to \mu^+ N \to \mu^+\mu^+ d \bar{u}$ and their charge conjugates will testify for the lepton number violation and a Majorana type $N$.

 \section{Model with multiple sterile neutrinos \label{multiple}}
In this section, we shall discuss the phenomenological consequences of a variant of the model introduced in sect.~\ref{phen} with more than one right-handed neutrinos with couplings of form
 \be Y_{\alpha i}\bar{N}_i\Phi^TcL_\alpha+{\rm H.c.}\label{Ni}\ee
 We shall take $Y_{\alpha i}\sim 3$. As we discussed in  sect.~\ref{phen}, adding multiple $N_i$ is motivated by the $(g-2)_\mu$ anomaly. We shall take all $N_i$ heavier than $\sim 3$ GeV to avoid the bounds from early universe, core collapse supernova and meson decay as well as from lower energy neutrino scattering experiments such a NOMAD. Integrating out the heavy $\Phi$ states, the coupling in Eq.~(\ref{Ni}) yields
 \be G_{ij}^\mu (\bar{N}_i\mu)(\bar{\mu}N_j) +G_{ij}^\nu (\bar{N}_i\nu_\mu)(\bar{\nu}_{\mu}N_j)\ee in which
 \be 
  G_{ij}^\mu=
 \frac{Y_{\mu i} Y_{\mu j}^*}{m_{\Phi^+}^2} \ \ \ {\rm and}\ \ \ G_{ij}^\nu=
 \frac{Y_{\mu i} Y_{\mu j}^*}{m_{\Phi^0}^2} +{\rm H.c}. \ee
 Taking $Y_{\mu i}\sim Y_{\mu j}\sim 3$ and $m_{\Phi^+}\sim m_{\Phi^0}\sim 300$~GeV, we find
 $G_{ij}^\nu\sim G_{ij}^\mu\sim 10^{-4}~{\rm GeV}^{-2}$. 
  All $N_i$ can be produced via the $\nu_\mu$ interaction in the forward experiments as described in the previous section. They can also be produced via the decay of the $\Phi$ components
at the Interaction Point of the LHC. The lightest $N_i$ decays into $\nu$+two jets or $\mu$+two jets as described  for the minimal model in the previous section.  
The heavier $N_i$ will however dominantly decay into lighter $N_j$ because $Y_{\mu i}\gg Y_u,Y_d$: 
$$ N_i \to N_j \mu\bar{\mu}  \ \ \ {\rm and} \ \ \ N_i \to N_j \nu_\mu\bar{\nu}_\mu.$$
For $m_{N_j}^2\ll m_{N_i}^2$, the decay length will be given by Eq.~(\ref{decayLENGHT}), replacing $|G_u|^2+|G_d|^2+|G_L|^2+|G_R|^2$ with $\sum_j(|G_{ij}^\mu|^2+|G_{ij}^\nu|^2)/3$ where $j$ includes all $N_j$ states lighter than $N_i$.  The factor of 3 is due to the summation on color in hadronic decay case. The factor $\sum_j(|G_{ij}^\mu|^2+|G_{ij}^\nu|^2)/3$ can be $\sim 30$ times larger than $|G_u|^2+|G_d|^2+|G_L|^2+|G_R|^2$, making the decay length 30 times smaller.  The whole decay chain of $N_i$ will take place inside the detector. As long as $N_i$ is lighter than $8-9$~GeV, the displacement of the vertex can still be resolved at FASER$\nu$. 

In case of $N_i \to \mu^-\mu^+ N_j$, the di-muon can be resolved at FASER$\nu$, providing a background-free signal even without resolving the displacement of the vertex \cite{Bakhti:2020szu}.  For $N_i \to \nu \bar{\nu} N_j$, the intermediate vertices in the decay chain of $N_i$ particles cannot however be identified and located. The decay vertex of the final $N$ is however guaranteed to be resolved  because the final vertex involves two detectable jets. If the lightest $N_i$ decays into muon+two jets, the total momenta of final particles can be measured. If there is a missing transverse momentum in the plane perpendicular to the line connecting the first neutral current type vertex and the second $\mu+$two jets vertex, this would indicate that the final $N$ was produced via decay chain in which $\nu\bar{\nu}$ pair(s)
were emitted rather than directly by interaction of $\nu_\mu$ from IP on the Tungsten at the detector.

\section{Model with a light scalar singlet \label{light-singlet}}
Within the model described in the previous sections, the $N$ particles are produced via $G_u$ and $G_d$ couplings which are suppressed by $m_{\Phi^0}^{-2}$. If we build a model in which the neutral mediator is lighter, the statistics can be higher. Let us consider a new singlet scalar $S$ which has a mixing angle $\theta$ with $\Phi^0$. Such a mixing may originate from trilinear coupling of form $AS^\dagger H^\dagger \cdot \Phi$. To preserve the global symmetry introduced in Eq.~(\ref{glob}), the relevant $U(1)$ charge of $S$ has to be equal to that of $\Phi$.  The cross section of $N$ production shown in Eq.~(\ref{diff-cross})  will then be modified with replacement
\be \label{rep} \left( \frac{m_{\Phi^0}^2}{t-m_{\Phi^0}^2}\right)^2\rightarrow  \left( \frac{m_{\Phi^0}^2\cos^2\theta }{t-m_{\Phi^0}^2}+ \frac{m_{\Phi^0}^2\sin^2\theta}{t-m_{S}^2}\right)^2.\ee
If $S$ is heavier than a few GeV, the bounds from meson decay can be avoided. Notice that $m_{\Phi^0}^2 \sin^2 \alpha$ should not be much larger than $m_{S}^2$; otherwise, the model will suffer from fine tuning. Taking $|t|<m_{S}^2< m_{\Phi^0}^2$ and a sizable mixing, the replacement  as  in Eq~(\ref{rep}) 
enhances the $N$ production cross section and therefore the statistics by a factor of 
$$4\times 10^2\left(\frac{30~{\rm GeV}}{m_{S}}\right)^2\frac{\sin^2\theta}{0.2}.$$
 In sect.~\ref{phen}, we  have observed that in the minimal version of the model,  the number of signal events cannot exceed $\sim 200$  for $m_N>3$~GeV. A number of signal events above $O(4000)$ would indicate a light singlet scalar mixed with $\Phi^0$. Such a scalar can be also produced at the Interaction Point of the LHC via mixing with $\Phi^0$. The produced $S$ decays into $N$ and $\nu_\mu$ with signatures that were already discussed in sect.~\ref{phen}.

Let us discuss the possibility of coherent enhancement of the cross section. As argued before $|t|\sim (10~{\rm GeV})^2(1-\cos \theta)$. In order for the amplitudes of the scatterings off various nucleons of Tungsten  to sum up coherently, $|t|$ should be of order of $(0.1~{\rm GeV})^2$ which corresponds to $(1-\cos\theta)\stackrel{<}{\sim}10^{-4}$.
Due to coherence, an enhancement of $A_W=183$ (corresponding to the mass number of Tungsten) is expected. As long as $m_S\gg 0.1$~GeV, no enhancement in the amplitude in this region is expected so the contribution of the coherent scattering with $|t|<(0.1~{\rm GeV})^2$ to the total cross section of scattering off Tungsten nucleus will be of order $A_W \Delta \cos\theta\sim 10^{-2}$ and therefore negligible.

\section{Summary and discussion\label{summary}}
We have proposed a model in which the left-handed doublet, $L_\mu=(\nu_\mu,\ \mu_L)$ couples to new scalar doublet(s), $\Phi=(\Phi^+, \ \Phi^0)$ and right-handed neutrino(s), $N$. If the components of $\Phi$ are heavier than $\sim 300$~GeV and their coupling to quarks is of order of $O(0.3)$ or smaller, they can escape the bounds from direct production at the colliders as well as the bounds from precision data, such as oblique parameters. Satisfying these bounds, the effective coupling between $L_\mu$ and quarks after integrating out the heavy $\Phi$ components can be as large as $10^{-5}$ GeV$^{-2}$.
In the minimal version of the model with only a single $N$, the signatures of the model at forward experiments will be a multiple jet vertex due to $N$ production (similar to that appearing due to the neutral current interaction of neutrinos in the SM), followed by a displaced vertex within a cone with an apex at the first vertex aligned in the direction of the beam with  a small opening of size $10^{-2}$. The topology of the event is shown in Fig.~\ref{fig:topology}.
The decay of $N$ can produce either two jets plus a charged muon $(u \bar{d}\mu)$ or two jets plus muon neutrino ($u \bar{u}\nu_\mu$ or $d\bar{d} \nu_\mu$). In the former case, all the final particles are detectable so the mass of $N$ can be reconstructed  at forward experiments by measuring the four momenta of the final particles. In the case of $N\to u \bar{u}\nu_\mu, d \bar{d}\nu_\mu$, we have formulated the condition under which the emission of $\nu_\mu$ can be observationally distinguished by measurement of the transverse momenta of the final jets. 

Notice that $\Phi$ and $N$ can also couple to the first and third generations of left-handed leptons, $L_e$ and $L_\tau$. We have however focused on the second generation as it is less constrained than the first generation. Moreover, with a coupling to the second generation, the possibility of an observable signal at forward experiments is higher thanks to the larger fluxes of $\nu_\mu$ and $\bar{\nu}_\mu$ compared to those of $\nu_e$, $\bar{\nu}_e$, $\nu_\tau$ and $\bar{\nu}_\tau$.

We have shown that the minimal model with only single $N$ and $\Phi$ can at most account for 25 \% of the $(g-2)_\mu$ anomaly but by adding more generations of $\Phi$ and/or $N$, the anomaly can be completely explained. In this case, the heavier $N$ will go through chain decays to the lightest $N$, emitting either $\mu^-\mu^+$ pairs or $\nu_\mu \bar{\nu}_\mu$ pairs in the process. The final $N$ will decay into jets plus a lepton, similarly to the case of single $N$. The $\mu^-\mu^+$ pair emitted through chain decay can be of course observed at forward experiments. We have shown that even in case that the decays of  intermediate $N$ particles produce only invisible $\nu_\mu \bar{\nu}_\mu$, their production can be confirmed by measuring the transverse momentum of the final particles.

We have argued that in both minimal version of the model with a single $N$
and in its variants with multiple $N$, the predicted signals at FASER$\nu$ will be background free. As a result, even a single event observed at forward experiments can count as discovery. We have shown that in the case of null results at FASER$\nu$ or SND@LHC, the bound on the relevant effective  couplings can be lowered down to $10^{-6}$ GeV$^{-2}$.
If $N$ is heavier than $O(3)$ GeV, it could not be produced at lower energy scattering experiments such as NOMAD \cite{NOMAD:2001xxt}, CHORUS \cite{CHORUS:2007wlo}, Miner$\nu$a \cite{MINERvA:2021csy}, CHARM II \cite{CHARMII:2008qag} and MicroBooNE \cite{MicroBooNE:2021fdt}.
For the same reason, the future state-of-the-art DUNE experiment cannot test the model, either. The energy of the neutrino beam at NuTeV experiment \cite{Dore:2018ldz} was few 100~GeV so $N$ particles heavier than 3~GeV could be produced at this experiment but the spatial resolution of NuTeV \cite{NuTeV:1999uck} was not fine enough to disentangle the second displaced vertex so the signals would be mistaken for SM NC and CC vertices. IF FASER$\nu$ discovers signals for $N$, the NuTeV data should be re-analyzed to correct parton distribution function derived from it.   Thus, the bound from FASER$\nu$ and SND@LHC will be the strongest, only to be surpassed by the bounds to be provided by their own upgrades.
The upgrade of FASER$\nu$ for the high luminosity phase of the LHC (FASER$\nu$2) can improve the bound down to $10^{-7}$ GeV$^{-2}$.

 If a discovery is made by forward experiments, it will be a strong motivation for customized searches for the $\Phi^+$ and $\Phi^0$ production at the CMS and ATLAS. These particles can be pair produced via electroweak interactions or can be singly produced in association of a gluon via their Yukawa couplings to the quarks at Interaction Point. They will then go through decays as $\Phi^+ \to N\mu^+ \to {\rm two ~ jets}+\mu^+\mu^-$ or $\to{\rm two ~ jets}+\mu^+\nu_\mu$ and 
$\Phi^+ \to N\bar{\nu}_\mu \to {\rm two ~ jets}+\mu^-\nu_\mu$ or $\to {\rm two ~ jets}+\bar{\nu}_\mu\nu_\mu$. Moreover, if $N$ is a Majorana particle, it can decay into $\mu^+$ instead of $\mu^-$, producing a same sign muon signal. Thus, the Majorana nature of $N$ can be established by CMS and ATLAS via detecting  same sign muon signals. 

If $N$ is of Majorana type, its couplings to $\nu_\mu$ can contribute to the $\mu\mu$ component of neutrino mass matrix. Then, the smallness of neutrino mass imposes a bound on the mass splittings between the   CP-even and CP-odd components of $\Phi^0$. We have devised a global U(1) symmetry explaining this smallness. The same symmetry can also explain the smallness of the masses of the first generation quarks as a bonus.

The $N$ particles can also be produced by high energy atmospheric neutrinos scattering off nuclei inside the neutrino telescopes. The production of $N$ will lead to a cascade similar to those produced by the SM neutral current interactions. Since the cross section of the new interactions is at most 10 \% of the cross section of the SM neutral current, the new physics can account for less than $10\%$ of the cascade events registered by ICECUBE. The produced $N$ will decay after traveling $\sim 2~{\rm m}(3~{\rm GeV}/m_N)^6 (E_N/100~{\rm TeV})$. For heavy $N$, the decay length will be too short to be resolved by ICECUBE. Moreover, the particles  from the $N$ decay will be emitted almost parallel, making a small angle of $m_N/(2E_N)$ with each other. Since their total electric charge is zero, it will be like propagation of a neutral particle in ice so the Cherenkov emission from the particles of the $N$ decay may be too faint to be detected at the neutrino telescopes.

%\appendix
%\section{Some title}
%Please always give a title also for appendices.

\acknowledgments

 This project has received funding /support from the European Union$^\prime$s Horizon 2020 research and innovation programme under the Marie Sklodowska -Curie grant agreement No 860881-HIDDeN. YF has received  financial support from Saramadan under contract No.~ISEF/M/400279 and No.~ISEF/M/99169.
SA is supported by a grant from Basic Sciences Research Fund (No. BSRF-phys-399-01).
%\paragraph{Note added.} This is also a good position for notes added
%after the paper has been written.
\bibliographystyle{JHEP}
\bibliography{Revision.bib}

\providecommand{\href}[2]{#2}\begingroup\raggedright\begin{thebibliography}{10}

\bibitem{NOMAD:2001xxt}
{\scshape NOMAD} collaboration, \emph{{Final NOMAD results on muon-neutrino
  ---\ensuremath{>} tau-neutrino and electron-neutrino ---\ensuremath{>}
  tau-neutrino oscillations including a new search for tau-neutrino appearance
  using hadronic tau decays}},
  \href{https://doi.org/10.1016/S0550-3213(01)00339-X}{\emph{Nucl. Phys. B}
  {\bfseries 611} (2001) 3}
  [\href{https://arxiv.org/abs/hep-ex/0106102}{{\ttfamily hep-ex/0106102}}].

\bibitem{Abreu:2019yak}
{\scshape FASER} collaboration, \emph{{Detecting and Studying High-Energy
  Collider Neutrinos with FASER at the LHC}},
  \href{https://doi.org/10.1140/epjc/s10052-020-7631-5}{\emph{Eur. Phys. J. C}
  {\bfseries 80} (2020) 61} [\href{https://arxiv.org/abs/1908.02310}{{\ttfamily
  1908.02310}}].

\bibitem{SND}
{\scshape SHiP} collaboration, \emph{{SND@LHC}},
  \href{https://arxiv.org/abs/2002.08722}{{\ttfamily 2002.08722}}.

\bibitem{Bakhti:2020vfq}
P.~Bakhti, Y.~Farzan and S.~Pascoli, \emph{{Unravelling the richness of dark
  sector by FASER$\nu$}},
  \href{https://doi.org/10.1007/JHEP10(2020)008}{\emph{JHEP} {\bfseries 10}
  (2020) 008} [\href{https://arxiv.org/abs/2006.05437}{{\ttfamily
  2006.05437}}].

\bibitem{Falkowski:2021bkq}
A.~Falkowski, M.~Gonz\'alez-Alonso, J.~Kopp, Y.~Soreq and Z.~Tabrizi,
  \emph{{EFT at FASER$\nu$}},
  \href{https://arxiv.org/abs/2105.12136}{{\ttfamily 2105.12136}}.

\bibitem{Ismail:2020yqc}
A.~Ismail, R.~Mammen~Abraham and F.~Kling, \emph{{Neutral current neutrino
  interactions at FASER$\nu$}},
  \href{https://doi.org/10.1103/PhysRevD.103.056014}{\emph{Phys. Rev. D}
  {\bfseries 103} (2021) 056014}
  [\href{https://arxiv.org/abs/2012.10500}{{\ttfamily 2012.10500}}].

\bibitem{Kling:2020iar}
F.~Kling, \emph{{Probing light gauge bosons in tau neutrino experiments}},
  \href{https://doi.org/10.1103/PhysRevD.102.015007}{\emph{Phys. Rev. D}
  {\bfseries 102} (2020) 015007}
  [\href{https://arxiv.org/abs/2005.03594}{{\ttfamily 2005.03594}}].

\bibitem{Bakhti:2020szu}
P.~Bakhti, Y.~Farzan and S.~Pascoli, \emph{{Discovery potential of FASER$\nu$
  with contained vertex and through-going events}},
  \href{https://doi.org/10.1007/JHEP04(2021)075}{\emph{JHEP} {\bfseries 04}
  (2021) 075} [\href{https://arxiv.org/abs/2010.16312}{{\ttfamily
  2010.16312}}].

\bibitem{Jho:2020jfz}
Y.~Jho, J.~Kim, P.~Ko and S.C.~Park, \emph{{Search for sterile neutrino with
  light gauge interactions: recasting collider, beam-dump, and neutrino
  telescope searches}},  \href{https://arxiv.org/abs/2008.12598}{{\ttfamily
  2008.12598}}.

\bibitem{Jodlowski:2020vhr}
K.~Jod\l{}owski and S.~Trojanowski, \emph{{Neutrino beam-dump experiment with
  FASER at the LHC}},
  \href{https://doi.org/10.1007/JHEP05(2021)191}{\emph{JHEP} {\bfseries 05}
  (2021) 191} [\href{https://arxiv.org/abs/2011.04751}{{\ttfamily
  2011.04751}}].

\bibitem{Ismail:2021dyp}
A.~Ismail, S.~Jana and R.M.~Abraham, \emph{{Neutrino Up-scattering via the
  Dipole Portal at Forward LHC Detectors}},
  \href{https://arxiv.org/abs/2109.05032}{{\ttfamily 2109.05032}}.

\bibitem{vonBuddenbrock:2017gvy}
S.~von Buddenbrock, A.S.~Cornell, A.~Fadol, M.~Kumar, B.~Mellado and X.~Ruan,
  \emph{{Multi-lepton signatures of additional scalar bosons beyond the
  Standard Model at the LHC}},
  \href{https://doi.org/10.1088/1361-6471/aae3d6}{\emph{J. Phys. G} {\bfseries
  45} (2018) 115003} [\href{https://arxiv.org/abs/1711.07874}{{\ttfamily
  1711.07874}}].

\bibitem{Fischer:2021sqw}
O.~Fischer et~al., \emph{{Unveiling Hidden Physics at the LHC}},  in
  \emph{{Unveiling hidden Physics Beyond the Standard Model at the LHC}}, 9,
  2021 [\href{https://arxiv.org/abs/2109.06065}{{\ttfamily 2109.06065}}].

\bibitem{Hernandez:2019geu}
Y.~Hernandez, M.~Kumar, A.S.~Cornell, S.-E.~Dahbi, Y.~Fang, B.~Lieberman
  et~al., \emph{{The anomalous production of multi-lepton and its impact on the
  measurement of $Wh$ production at the LHC}},
  \href{https://doi.org/10.1140/epjc/s10052-021-09137-1}{\emph{Eur. Phys. J. C}
  {\bfseries 81} (2021) 365}
  [\href{https://arxiv.org/abs/1912.00699}{{\ttfamily 1912.00699}}].

\bibitem{Sabatta:2019nfg}
D.~Sabatta, A.S.~Cornell, A.~Goyal, M.~Kumar, B.~Mellado and X.~Ruan,
  \emph{{Connecting muon anomalous magnetic moment and multi-lepton anomalies
  at LHC}}, \href{https://doi.org/10.1088/1674-1137/44/6/063103}{\emph{Chin.
  Phys. C} {\bfseries 44} (2020) 063103}
  [\href{https://arxiv.org/abs/1909.03969}{{\ttfamily 1909.03969}}].

\bibitem{Mathaha:2021buc}
T.~Mathaha, A.K.~Swain, M.~Kumar, X.~Ruan and B.~Mellado, \emph{{Understanding
  two same-sign and three leptons with $b$-jets in four top quark events at the
  LHC}},  in \emph{{65th Annual Conference of the South African Institute of
  Physics}}, 9, 2021 [\href{https://arxiv.org/abs/2109.06951}{{\ttfamily
  2109.06951}}].

\bibitem{Crivellin:2021ubm}
A.~Crivellin, Y.~Fang, O.~Fischer, A.~Kumar, M.~Kumar, E.~Malwa et~al.,
  \emph{{Accumulating Evidence for the Associate Production of a Neutral Scalar
  with Mass around 151 GeV}},
  \href{https://arxiv.org/abs/2109.02650}{{\ttfamily 2109.02650}}.

\bibitem{Fowlie:2021ldv}
A.~Fowlie, \emph{{Comment on ''Accumulating Evidence for the Associate
  Production of a Neutral Scalar with Mass around 151 GeV''}},
  \href{https://arxiv.org/abs/2109.13426}{{\ttfamily 2109.13426}}.

\bibitem{Kling:2021gos}
F.~Kling, \emph{{Forward Neutrino Fluxes at the LHC}},
  \href{https://arxiv.org/abs/2105.08270}{{\ttfamily 2105.08270}}.

\bibitem{felix_kling_2020_4059893}
F.~Kling and J.L.~Feng, \emph{Forward Physics Facility}, Aug., 2020.
\newblock 10.5281/zenodo.4059893.

\bibitem{Haller:2018nnx}
J.~Haller, A.~Hoecker, R.~Kogler, K.~M\"onig, T.~Peiffer and J.~Stelzer,
  \emph{{Update of the global electroweak fit and constraints on
  two-Higgs-doublet models}},
  \href{https://doi.org/10.1140/epjc/s10052-018-6131-3}{\emph{Eur. Phys. J. C}
  {\bfseries 78} (2018) 675}
  [\href{https://arxiv.org/abs/1803.01853}{{\ttfamily 1803.01853}}].

\bibitem{Allwicher:2021rtd}
L.~Allwicher, P.~Arnan, D.~Barducci and M.~Nardecchia, \emph{{Perturbative
  unitarity constraints on generic Yukawa interactions}},
  \href{https://arxiv.org/abs/2108.00013}{{\ttfamily 2108.00013}}.

\bibitem{buckley2015lhapdf6}
A.~Buckley, J.~Ferrando, S.~Lloyd, K.~Nordstr\"om, B.~Page, M.~R\"ufenacht
  et~al., \emph{{LHAPDF6: parton density access in the LHC precision era}},
  \href{https://doi.org/10.1140/epjc/s10052-015-3318-8}{\emph{Eur. Phys. J. C}
  {\bfseries 75} (2015) 132} [\href{https://arxiv.org/abs/1412.7420}{{\ttfamily
  1412.7420}}].

\bibitem{ball2017parton}
{\scshape NNPDF} collaboration, \emph{{Parton distributions from high-precision
  collider data}},
  \href{https://doi.org/10.1140/epjc/s10052-017-5199-5}{\emph{Eur. Phys. J. C}
  {\bfseries 77} (2017) 663}
  [\href{https://arxiv.org/abs/1706.00428}{{\ttfamily 1706.00428}}].

\bibitem{Lavoura:2003xp}
L.~Lavoura, \emph{{General formulae for f(1) ---\ensuremath{>} f(2) gamma}},
  \href{https://doi.org/10.1140/epjc/s2003-01212-7}{\emph{Eur. Phys. J. C}
  {\bfseries 29} (2003) 191}
  [\href{https://arxiv.org/abs/hep-ph/0302221}{{\ttfamily hep-ph/0302221}}].

\bibitem{Farzan:2009ji}
Y.~Farzan, \emph{{A Minimal model linking two great mysteries: neutrino mass
  and dark matter}},
  \href{https://doi.org/10.1103/PhysRevD.80.073009}{\emph{Phys. Rev. D}
  {\bfseries 80} (2009) 073009}
  [\href{https://arxiv.org/abs/0908.3729}{{\ttfamily 0908.3729}}].

\bibitem{Muong-2:2021ojo}
{\scshape Muon g-2} collaboration, \emph{{Measurement of the Positive Muon
  Anomalous Magnetic Moment to 0.46 ppm}},
  \href{https://doi.org/10.1103/PhysRevLett.126.141801}{\emph{Phys. Rev. Lett.}
  {\bfseries 126} (2021) 141801}
  [\href{https://arxiv.org/abs/2104.03281}{{\ttfamily 2104.03281}}].

\bibitem{Boehm:2006mi}
C.~Boehm, Y.~Farzan, T.~Hambye, S.~Palomares-Ruiz and S.~Pascoli, \emph{{Is it
  possible to explain neutrino masses with scalar dark matter?}},
  \href{https://doi.org/10.1103/PhysRevD.77.043516}{\emph{Phys. Rev. D}
  {\bfseries 77} (2008) 043516}
  [\href{https://arxiv.org/abs/hep-ph/0612228}{{\ttfamily hep-ph/0612228}}].

\bibitem{Farzan:2010fw}
Y.~Farzan and M.~Hashemi, \emph{{SLIM at LHC: LHC search power for a model
  linking dark matter and neutrino mass}},
  \href{https://doi.org/10.1007/JHEP11(2010)029}{\emph{JHEP} {\bfseries 11}
  (2010) 029} [\href{https://arxiv.org/abs/1009.0829}{{\ttfamily 1009.0829}}].

\bibitem{CHORUS:2007wlo}
{\scshape CHORUS} collaboration, \emph{{Final results on nu(mu)
  ---\ensuremath{>} nu(tau) oscillation from the CHORUS experiment}},
  \href{https://doi.org/10.1016/j.nuclphysb.2007.10.023}{\emph{Nucl. Phys. B}
  {\bfseries 793} (2008) 326}
  [\href{https://arxiv.org/abs/0710.3361}{{\ttfamily 0710.3361}}].

\bibitem{MINERvA:2021csy}
{\scshape MINERvA} collaboration, \emph{{Exploring Neutrino-Nucleus
  Interactions in the GeV Regime using MINERvA}},
  \href{https://arxiv.org/abs/2107.02064}{{\ttfamily 2107.02064}}.

\bibitem{CHARMII:2008qag}
{\scshape CHARM II} collaboration, \emph{{Neutral current coupling constants
  from $\nu_{\mu}e^-$and $\bar{\nu}_{\mu}e^-$ scattering}},
  \href{https://doi.org/10.1063/1.43528}{\emph{AIP Conf. Proc.} {\bfseries 272}
  (2008) 727}.

\bibitem{MicroBooNE:2021fdt}
{\scshape MicroBooNE} collaboration, \emph{{First Measurement of Inclusive
  Electron-Neutrino and Antineutrino Charged Current Differential Cross
  Sections in Charged Lepton Energy on Argon in MicroBooNE}},
  \href{https://arxiv.org/abs/2109.06832}{{\ttfamily 2109.06832}}.

\bibitem{Dore:2018ldz}
U.~Dore, P.~Loverre and L.~Ludovici, \emph{{History of accelerator neutrino
  beams}}, \href{https://doi.org/10.1140/epjh/e2019-90032-x}{\emph{Eur. Phys.
  J. H} {\bfseries 44} (2019) 271}
  [\href{https://arxiv.org/abs/1805.01373}{{\ttfamily 1805.01373}}].

\bibitem{NuTeV:1999uck}
{\scshape NuTeV} collaboration, \emph{{Precision calibration of the NuTeV
  calorimeter}},
  \href{https://doi.org/10.1016/S0168-9002(99)01304-2}{\emph{Nucl. Instrum.
  Meth. A} {\bfseries 447} (2000) 377}
  [\href{https://arxiv.org/abs/hep-ex/9908056}{{\ttfamily hep-ex/9908056}}].

\end{thebibliography}\endgroup


\end{document}